\documentclass[conference]{IEEEtran}
\IEEEoverridecommandlockouts
\usepackage{cite}
\usepackage{amsmath,amssymb,amsfonts}
\usepackage{algorithmic}
\usepackage{graphicx}
\usepackage{textcomp}
\usepackage{xcolor}

\usepackage{algorithm}
\usepackage{algorithmic}
\usepackage{bm}
\usepackage{subcaption}
\usepackage{booktabs}
\usepackage{amssymb}
\usepackage{color}

\def\Black{\color{black}}

\def\ignore#1{}

\begin{document}

\title{Reducing Redundant Computation in Multi-Agent Coordination through Locally Centralized Execution  
}

\author{\IEEEauthorblockN{Yidong Bai}
  \IEEEauthorblockA{\textit{Dept. Computer Science and Engineering}\\
    \textit{Waseda University}\\
    \textit{Tokyo, Japan}\\
    ydbai@moegi.waseda.jp}
\and
\IEEEauthorblockN{Toshiharu Sugawara}
  \IEEEauthorblockA{\textit{Dept. Computer Science and Engineering}\\
    \textit{Waseda University}\\
    \textit{Tokyo, Japan}\\
    sugawara@waseda.jp}
}

\maketitle

\begin{abstract}
In \emph{multi-agent reinforcement learning}, decentralized execution is a common approach, yet it suffers from the \emph{redundant computation} problem. 
This occurs when multiple agents redundantly perform the same or similar computation due to overlapping observations. 
To address this issue, this study introduces a novel method referred to as \emph{locally centralized team transformer} (LCTT). 
LCTT establishes a \emph{locally centralized execution} framework where selected agents serve as \emph{leaders}, issuing \emph{instructions}, while the rest agents, designated as \emph{workers}, act as these instructions without activating their policy networks.
For LCTT, we proposed 
the \emph{team-transformer} (T-Trans) architecture that allows leaders to provide specific instructions to each worker, and the \emph{leadership shift} mechanism that allows agents autonomously decide their roles as leaders or workers. 
Our experimental results demonstrate that the proposed method effectively reduces redundant computation, does not decrease reward levels, and leads to faster learning convergence.
\end{abstract}


\section{Introduction}
Multi-agent deep reinforcement learning (MADRL) has made remarkable 
advancements recently~\cite{yu2019distributed,berner2019dota}. 
Two frameworks, \emph{centralized training and decentralized execution}
(CTDE)~\cite{lowe2017multi}
and 
\emph{centralized training and centralized execution}
(CTCE)~\cite{gupta2017cooperative}, are usually employed for MADRL.
In the CTCE model, a centralized network processes the information from 
all agents and determines their joint actions. 
However, CTCE is practical only in limited scenarios because of the
complex interference of agents' actions and the exponential growth of
the joint action space with the number of agents.
By contrast, agents in CTDE learn their policies using a 
centralized network, yet they execute decisions based on their 
individual observations.
This makes the CTDE approach more suitable for real-world applications
of multi-agent
systems~\cite{wooldridge2009introduction,sugawara1990cooperative}.
\par

\begin{figure}[tb]
\centering
\includegraphics[width=0.72\columnwidth]{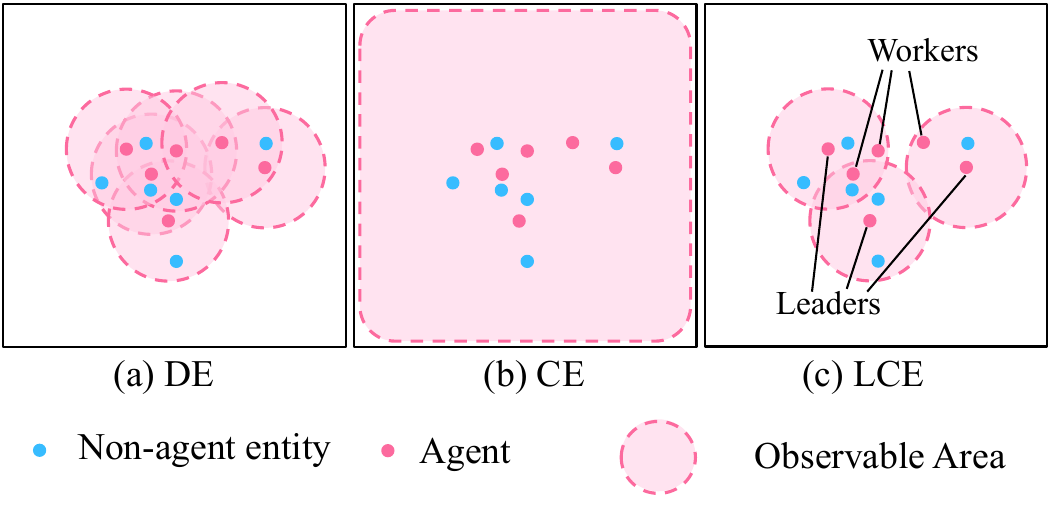}
\caption{Example Environments of MADRL.}
  \label{fig:Intro-LocCen}
\end{figure}

Yet several problems with CTDE frameworks have long been overlooked.
In particular, \emph{redundant computation} is a well-known traditional 
problem in multi-agent systems (MASs)~\cite{durfee1987coherent},
meaning that the similar computation is done
in some agents redundantly.
Such redundant computations are caused by overlapping
observations, resulting in the associated similar inference in different
agents. 
As depicted in Fig.~\ref{fig:Intro-LocCen}a,
agents in {\em decentralized execution} (DE), in which they observe and process
their local information, often have overlapping 
fields of observation.
This overlap means that information about various entities in the environment, 
like other agents, obstacles, and objectives, is processed repeatedly 
by several agents as they determine their next steps,
which seems as a waste of computing, especially 
in scenarios requiring teamwork and inter-agent communication.
By contrast, this issue of redundant computation is avoided in the
CTCE framework (Fig.~\ref{fig:Intro-LocCen}b), where all observed data
are summed up and provided to a central network.
\par

\def\redc{R_{\it dd}}

Appropriate mitigation of duplicate observations can significantly
reduce waste in calculations without compromising the effectiveness of
MASs. Thus, a hybrid execution framework called the \emph{locally
  centralized execution}~(LCE) as shown in
Fig.~\ref{fig:Intro-LocCen}c, in which agents perform less overlapped
observations and less redundant processes, is introduced.
In conjunction with \emph{centralized training}~(CT), the LCE
framework designates certain agents as \emph{leaders}. These leaders are
responsible for making decisions not only for themselves but also for
other agents, termed \emph{workers}, within their field of
view. Consequently, workers do not need to process their surroundings
or decide on their actions independently.
Unlike models that depend on aggregating all agents' observations, the
LCE approach cuts down on unnecessary computations and lowers the
complexity of decision-making by operating within a significantly
smaller joint action space.
\par

We then introduce a {\em centralized training and localized execution}
(CTLCE) framework based on LCE and propose a {\em localized centralized
team transformer} (LCTT) to create target messages to others.
Additionally, we also introduce the
\emph{redundant observation ratio} $\redc$ ($\geq1$) to measure the extent of redundant computation within an MAS algorithm, 
This ratio allows for a direct comparison of the computational expenses
between different deep learning-based methods, assuming they have a 
comparable number of parameters.
To our knowledge, This study is the \emph{first} attempt to address redundant
computational issue in MADRL.
\par

The proposed LCTT comprises of LCE, \emph{team transformer}~(T-Trans),
and \emph{leadership shift} (LS). LCE first establishes a cooperative
framework where agents are dynamically organized into leader and
worker roles, based on the directionality of instruction messages.
Then, T-Trans utilizes a mechnizm simitar to the
attention~\cite{vaswani2017attention} to allow leaders to dispatch
specific instructions to workers under its control.
Third, we employ the
\emph{leadership Q-values}, which are calculated for agents to
identify their aptness as leaders. Subsequently, we  propose LS, which enables
a current leader to decide which agents within their observation 
range are best suited to assume leadership in the following time step, 
based on their leadership Q-values. 
This process allows for a fluid transition of leadership roles among agents, ensuring that leadership is always assigned to the most qualified agents throughout the interaction.
\par

The proposed LCTT was implemented upon
QMIX~\cite{rashid2020monotonic} and then trained within the CTLCE
framework. We conducted its experimental evaluatation in a {\em
  level-based
  foraging} (LBF) problemt~\cite{yuan2022multi,papoudakis2020benchmarking},
by comparing the performance with the baselines, {\em multi-agent incentive
  communication}~(MAIC)~\cite{yuan2022multi} and QMIX without LCTT.
We shows that LCTT achieved faster convergence with comparable rewards
owing to the significant reduction of redundancy in computation.  
\par

\section{Related Work}
In addressing challenges within multi-agent learning, one
straightforward approach involves utilizing a centralized network that
determines joint actions to guide the behavior of all agents. This
approach is known as
CTCE~\cite{gronauer2022multi,han2019grid}. However, as the number of
agents increases, the joint
action space grows exponentially, which restricts the scalability of
CTCE to scenarios with many agents due to computational limitations.
Another scheme is DTDE~\cite{gronauer2022multi}, in which each agent
employs reinforcement learning individually to develop its
policy~\cite{tan1993multi}. By training policies independently, DTDE
sidesteps the issue of an exponentially expanding joint action space,
offering a more scalable solution for multi-agent environments.
\par

However, from the perspective of individual agents,
the challenge with DTDE lies in the non-stationary nature of the environment,
which changes dynamically due to the actions and policies of other
agents that also learn.
This often leads to difficulties in achieving convergence.
In contrast, CTDE methods~\cite{rashid2020monotonic,lowe2017multi} 
learn a shared policy
for all agents in a centralized manner while allowing agents to act based on
their own local observations.
Therefore, CTDE mitigates the problem of non-stationarity and reduces
the complexities associated with a huge joint action space.
However, DE in CTDE and DTDE encounters an issue of redundant
computations. Thus, we propose LCE to mitigate redundant computations
and also propose a CTLCE framework, which can be regarded as an
intermediate scheme of CTCE and CTDE.
\par

\def\AgentSet{\mathcal{N}}
\def\ActionSet{\mathcal{A}}
\def\StateSet{\mathcal{S}}
\def\Null{\mathit{null}}

\section{Background and Problem}
\subsection{Cooperative Dec-POMDP with Communication}
Similar to some conventional
methods~\cite{ding2020learning,wang2022tomc,jiang2018learning,yuan2022multi},
our model is based on a fully cooperative {\em decentralized partially
  observable Markov decision process} (Dec-POMDP) augmented with
communication.
Let $\StateSet$ be the set of global states and
$\AgentSet=\{1, \dots, n\}$ represent the set of $n$ agents.
We also denote $\ActionSet$ as the set of actions that can be executed
by $\forall i\in \AgentSet$.
Discrete time $t\geq0$ is introduced but
we often omit the the subscript if $t$ for the sake of simplicity.
At every time $t$, agent $i\in\AgentSet$ in state $s \in
\StateSet$ obtains a local observation $o_i = O (s,i)$.
Then agent $i$ may
send a message $m_i$ based on $o_i$ to the agents in the
observable area.
Concurrently, $i$ might also receive messages from other agents; these
messages are denoted as vector
$\bm{m_{*,i}}=(m_{1,i}, \dots, m_{n,i})$ whose element $m_{j,i}$ denotes the
message from $j$ to $i$ ($i,i\in\AgentSet$).
We set $m_{j,i}=\Null$ if message did not arrive from $j$.
Then, referring to $\bm{m_{*,i}}$ and 
$o_i$, $i$ selects $a_i \in \ActionSet$ using its policy $\pi_i(a_i|o_i,
\bm{m_{*,i}})$ and executes it. After that, $i$ might receive reward
$r_i(s,a_i)$ and the state transistions to the next state $s'\in\StateSet$.
Agent $i$ attempts to 
to increase the expected value of the discounted cumulative reward
$R_i=\mathbb{E}[\sum_{t}\gamma^tr_i(s,a_i)]$ as much as possible
through learnig of $\pi_i$.
\par

We employ DQN~\cite{mnih2015human} to learn Q-function,
$Q(o_i,a_i|\theta)$, that enables agents to appropriately decide their
actions, where $\theta$ is the parameters of the associated deep
neural network.
Parameters $\theta$ is adjusted through learned to minimize
the loss function $L^Q(\theta)$.
\begin{equation} \label{eq:loss-DQN}
  L^Q(\theta) = \mathbb{E}_{o_i,a_i,r_i,o'_i}
  [(r_i + 
  \gamma\textrm{max}_{a'_i}\bar{Q}(o'_i,a'_i|\bar{\theta}) -
  Q(o_i,a_i|\theta))^2], 
\end{equation}
where $\bar{Q}$ denotes the output from the target Q network parameterized by
$\bar{\theta}$ which is updated with
$\theta$ periodically,
and $a'_i$ and $o'_i$ are the action and observation at the
next time step,

\par

\def\food{f}
\def\foodlv{lv_f}
\def\agentlv{lv_a}

\subsection{Level-Based Foraging (LBF) Environment}
\begin{figure}[tb]
\centering
\includegraphics[width=0.33\columnwidth]{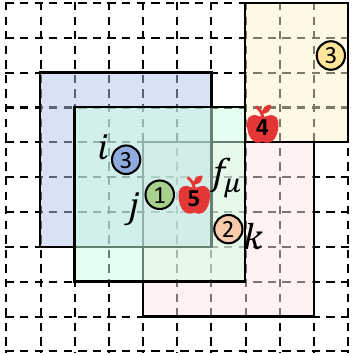}
\caption{Example of a level-based foraging environment.}
\label{fig:LBF}
\end{figure}

In a LBF environment, agents are tasked with collaboratively 
loading and collecting food. 
An example snapshot of the LBF is shown in Fig.~\ref{fig:LBF}.
At the start of the game, $n$ agents (circles in Fig.~\ref{fig:LBF})
and  $\beta$ ($>0$) foods $\{\food_1, \dots, \food_\beta\}$ (apples in
in Fig.~\ref{fig:LBF}) 
are randomly scattered in the environment. 
Food $\food_\mu$ is assigned an association level $\foodlv(\food_\mu)>0$, 
reflecting the challenge associated with acquiring that 
food $\food_\mu$.
Similarly, agent $i$ is characterized by an associated level, 
$\agentlv(i)$, which represents the $i$'s capability to forage for
food. These levels are expressed by the numbers on foods and agents in in
Fig.~\ref{fig:LBF}.
Agents can observe local areas shown by the lighter color
shade of each agent in Fig.~\ref{fig:LBF},
communicate with each other, and execute the selected actions every time.
This set of actions $\ActionSet$ includes moving in four different
directions,  a ``none'' action indicating inactivity, and a
``loading'' action  which is used when an agent to attempt to collect
food $\food_\mu$ on the adjacent node.
\par

If $\foodlv(\food_\mu) > \agentlv(i)$, agent $i$
is unable to collect $\food_\mu$ individually and fails this attempt.
For instance,
agent $j$ in Fig.~\ref{fig:LBF} whose level is \emph{one},
will fail if it tries to collect
food at level of \emph{five} on its own, hence $j$ has to wait for
other agents (e.g., $k$ and $i$) to come up with the food. 
Then, when  $\sum_{i\in\AgentSet_{\food\mu}} \agentlv(i) \geq
\foodlv(\food_\mu)$, i.e., the
sum of the levels of the agents that try to load $\food_\mu$
simultaneously is larger than or equal to $\foodlv(\food_\mu)$,
they can collect, divide, and load it. Note that
$\AgentSet_{\food\mu}$ represents the set of agents that executes ``loading''
$\food_\mu$ at its neighbor nodes, simultaneously.
\par

When agents load $\food_\mu$ successfully,
they receive the rewards by dividing $\foodlv(\food_\mu)$ 
by their levels. Therefore,
The rewards of $i \in \AgentSet_{\food\mu}$ is
\[
  r_i(s,a_i) = \foodlv(\food_\mu) \cdot\frac{\agentlv(i)}{\sum_{\forall j \in \AgentSet_{\food\mu}} \agentlv(j)},
\]
which means that the reward for food and the level of that food are the same.
Each agent $i$ aims to maximize both their {\em individual
rewards} $R_i$ and {\em team rewards} $R=\sum_{i\in\AgentSet} R_i$. 
While agents collaborate to maximize $R$, the pursuit of maximizing 
$R_i$ may not always align with team cooperation, especially if an additional agent joins the effort to collect food unnecessarily.
\par

\subsection{Problem Statementn}
Agents $i$, $j$, and $k$ in Fig.~\ref{fig:LBF} observe food $f_\mu$
and communicate to cooperatively to load it. The information in
$i$'s observations, containing information about $f_\mu$, $i$ and $j$,
is included in $j$'s observations. Likewise, the information for $k$'s
observations is also included in $j$'s 
observation. This results in the repeated processing of the same
information across the networks for these agents, leading to wasted
use of computing resources.
This issue becomes even more pronounced within the CTDE framework
where agents operate under shared network parameters.
\par

Therefore, instead of allowing $i$, $j$, and $k$ make decisions
separately, 
a more efficient approach would be enabling agent $j$,
using $j$'s observation and
network, to coordinate the actions of
$i$ and $k$ through communication. 
This method prevents the need for executing networks in agents 
$i$ and $k$ redundantly. 
We posit that such a mechanism can reduce the required computing cost
and enhance the overall efficiency of MADRL.
Our research introduces a framework in which agents autonomously
determine which agents should instruct or be instructed.
\par

\section{Proposed Method}
\subsection{Redundant Observation Ratio}
We propose a metric \emph{redundant observation ratio} $\redc$ for DE
to describe the extent of redundant computation in MASs.
This metric quantifies 
the average number of times the information pertaining to each entity
is processed across agents. It is defined as follows.
\begin{equation} \label{eq:Rdd}
  \redc = \frac{\sum_{i\in\AgentSet} U_i}{\sum_{e\in E} \delta(e)},
\end{equation}
where $E$ is the set of entities, such as
all targets, agents, and obstacles, in the environment 
and $U_i$ represents
the number of entities in $i$'s observation.
Function $\delta$ express the observability of entity $e\in E$, i.e.,
$\delta(e)=1$ if an agent observes $e$; otherwise, $\delta(e)=0$. 
\par

Hence, $\redc$ indicates the extent of
overlaps in their observations and thus the associated redundant
computations. For instance, $\redc$ are $2.75$, $1.17$, and $1.0$,
respectively, in Figs.~\ref{fig:Intro-LocCen}a, \ref{fig:Intro-LocCen}c, and
\ref{fig:Intro-LocCen}d.
Usually, $\redc \geq 1$ holds.
Because the observations of
all agents are merged and fed to the centralized
network in the CE framework (Fig.~\ref{fig:Intro-LocCen}b),
we can consider $\redc = 1$.
\par

\def\LeaderSet{\mathcal{L}}
\def\WorkerSet{\mathcal{W}}

\subsection{Locally Centralized Execution~(LCE)}
First, let $l$ ($\leq n$) agents be leaders and the other $n-l$ agents be
workers.

To mitigate redundant observations and computations, 
we enable leaders to construct instructions of
actions for workers in their observable region including itself.
The workers act as the leaders' instructions and thus can eliminate the need
to observe their surroundings (and their observations were excluded
from Eq.~\ref{eq:Rdd}).
$\WorkerSet$ and $\LeaderSet$ are the sets of workers and leaders,
respectively. 
The instruction message, $m_{i,k}$, from $i$ (leader) to $k$ (worker)
encapsulates the Q value of $k$'s possible actions, 
$Q_{i,k}(o_i,a_k|\theta)$ for $a_k\in\ActionSet$ and observation $o_i$.
\par

\begin{figure}[tb]
\centering
\includegraphics[width=0.99\linewidth]{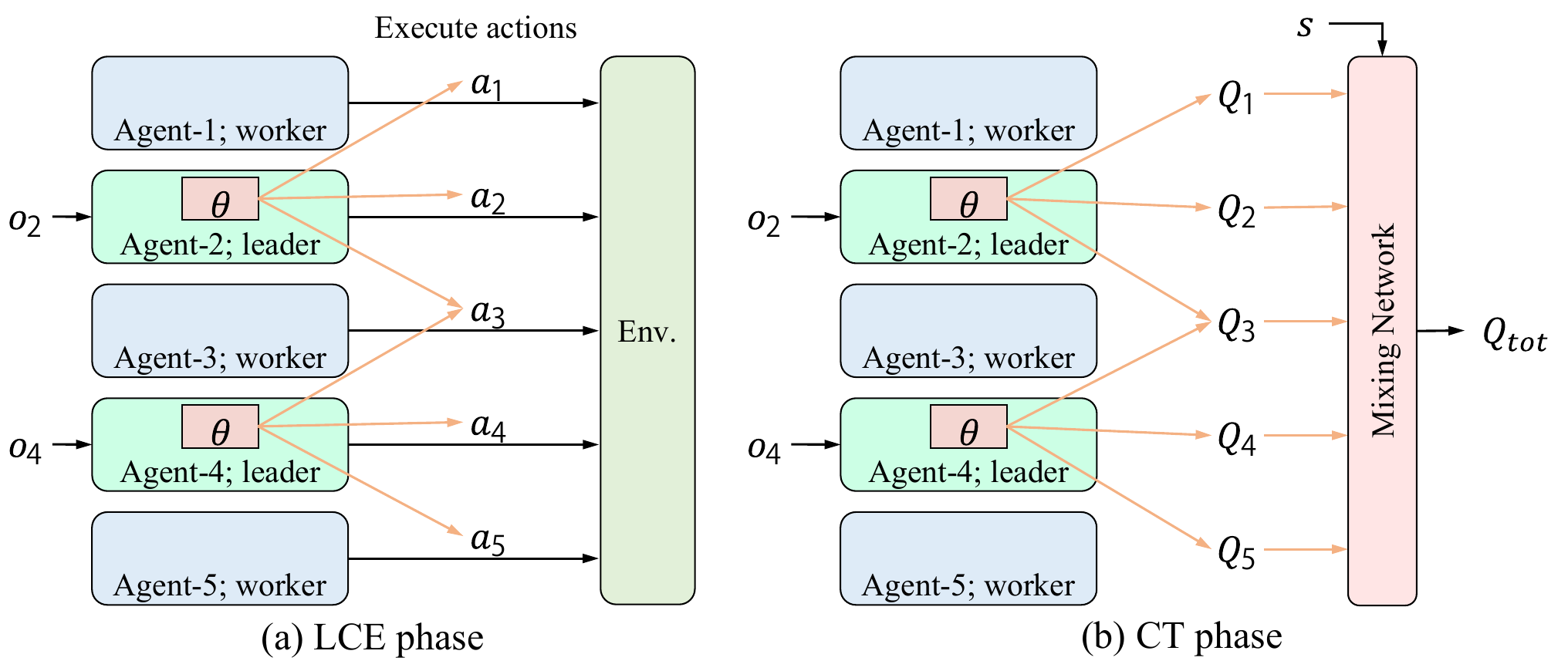}
\caption{Example of a CTLCE framework structure, in which
  $Q_{tot}$ is the global Q-values, $s$ is the global state, and
  $\theta$ is the policy network.}
\label{fig:CTLCE-basic}
\end{figure}

An example structure of the CTLCE
framework is shown in Fig.~\ref{fig:CTLCE-basic}, in which agents are
able to observe their neighboring two agents. 
Agents $2$ and $4$ are leaders and agents $1,3$ and $5$ are workers.
The actions of agents $1,3$ and $5$ are instructed by their adjacent
leaders in the LCE phase (Fig.~\ref{fig:CTLCE-basic}a). In CT phase
(Fig.~\ref{fig:CTLCE-basic}b), the policy is trained similar to
standard QMIX~\cite{rashid2020monotonic}.
\par

Algorithm~\ref{alg:LCE} shows the pseudocode for LCE.
If leaders are observable each other, they construct the instructions
for other leaders as well as for themselves.
When a agent $j$ receives multiple instruction from various leaders,
it averages these instructions to determine the Q values for its
actions.
In the LCE framework, the number of leaders is typically set to be
 $0<l<n$; LCE with $l=0$ represents
DE without any form of communication between agents,
and that with $l=n$ indicates DE where there is dense communication
across all agents. Finding an appropriate value for $l$ is crucial for
the efficiency of the framework and often requires experiments and
experience.
\par

All agents within the system share the same neural network
architecture, although workers typically do not activate their
networks as often as leaders do. 
Only when a worker does not receive any instruction,
the worker observes the surroundings and run its own network to obtain
Q-values for itself (Line~\ref{alg:LCE+worker1}$\sim$\ref{alg:LCE+worker2} 
in Alg.~\ref{alg:LCE}).
LS (Line~\ref{alg:LCE-LS} in Alg.~\ref{alg:LCE})
and how leaders constuct instructions~(Lines~\ref{alg:LCE-mij} \& 
\ref{alg:LCE-mkk} in Alg.~\ref{alg:LCE}) are
are explained in Sections~\ref{sec:TeamTransformer} and
\ref{sec:LeadershipShift}.
\par

\renewcommand{\algorithmicwhile}{\textbf{parallel for}}

\begin{algorithm}[tb]
\caption{Locally Centralized Execution}
\label{alg:LCE}
\begin{algorithmic}[1] 
\STATE Let $t=0$. We randomly designate $l$ agents as leaders,
and the other $(n-l)$ agents as workers. \label{alg:LCE-l-leaders}
\STATE
\FOR{$t=0, \dots, T$}
    \WHILE{All leaders $i \in \LeaderSet$}
        \STATE Obtain local observation $o_i$.
        \WHILE{Observed agents $j$ (including $i$)}
            \STATE Construct an instruction $m_{i,j}$ \label{alg:LCE-mij}
            \STATE Send $m_{i,j}$ to agent $j$
        \ENDWHILE
        \STATE Leadership Shift: Appoint an observed agent
        to be the leader at the $t+1$.\label{alg:LCE-LS}
    \ENDWHILE

    \WHILE{All worker $k \in \WorkerSet$} \label{alg:LCE+worker1}
        \IF{$k$ receives no instruction}
            \STATE Obtain local observation $o_k$.
            \STATE Construct an instruction for itself, $m_{k,k}$  \label{alg:LCE-mkk}
        \ENDIF 
    \ENDWHILE \label{alg:LCE+worker2}
    
    \STATE Calculate average values of instructions as Q-values
    for all agents $\AgentSet$
    \STATE Agents act as the Q-values 
\ENDFOR
\end{algorithmic}
\end{algorithm}

\begin{figure*}[tb]
\centering
\includegraphics[width=0.66\textwidth]{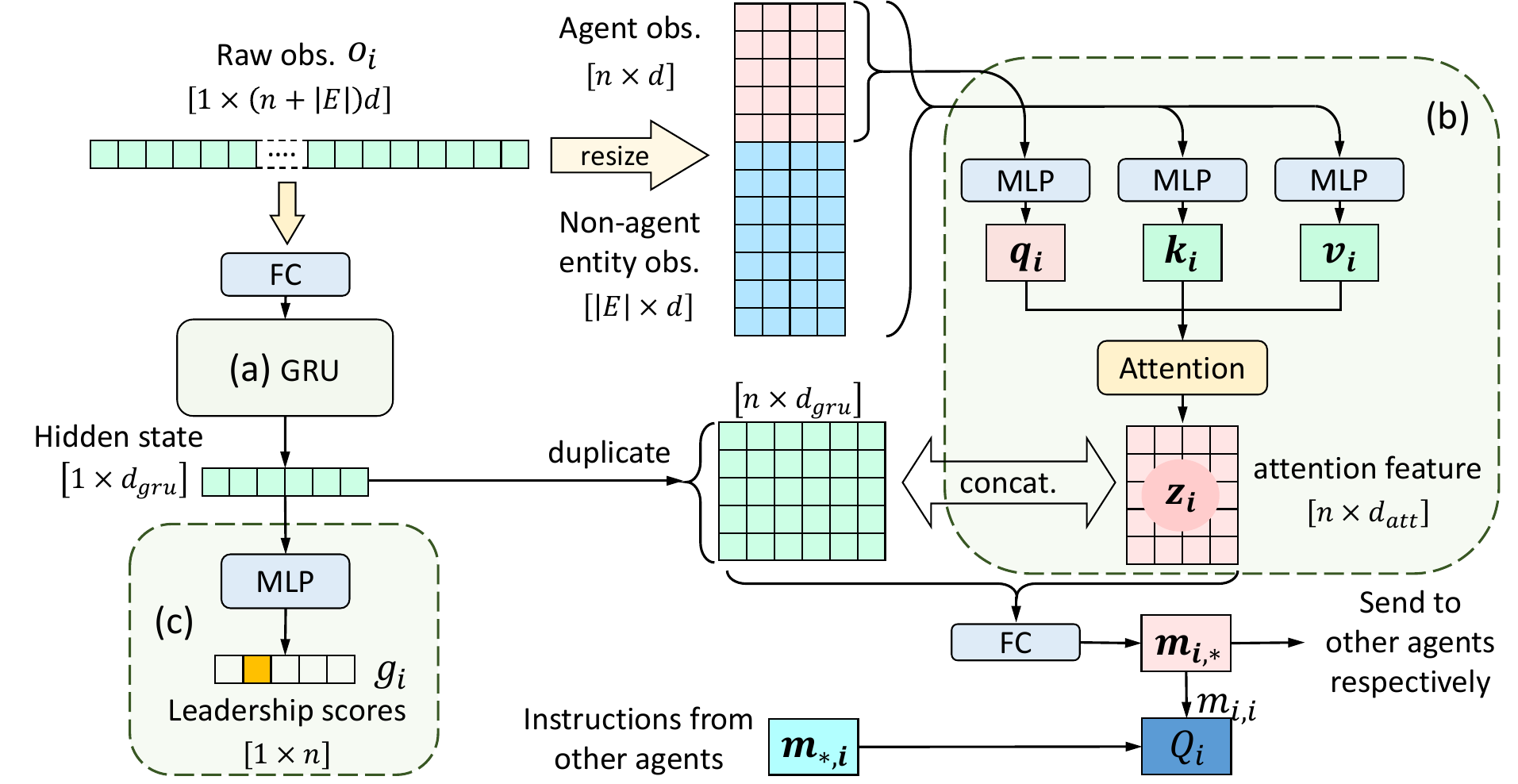}
\caption{T-Trans Structure. Parameters $d_{att}$ and $d_{gru}$
  are dimensions of features of attention layers and GRU, and $d$ is
  the dimension of observation of respective entities; 
$\bm{k_i}$ and $\bm{v_i}$  are key and value matrices whose sizes 
  are $(|E|+n) \times d_{att}$. 
$\bm{q_i}$ is the $n \times d_{att}$ query matrix.
}
\label{fig:Team-Trans}
\end{figure*}

\subsection{Team Transformer (T-Trans)}\label{sec:TeamTransformer}
In existing research, teammate
modeling~\cite{wang2022tomc,yuan2022multi} 
has been employed to allow agents to to send specific messages 
to a designated teammate. 
Unfortunately, this approach is unsuitable for our leaders,
as workers do not (always) determine their actions using
their own policies.
Therefore, we introduce T-Trans so that leaders are able to
generate targeted messages for their workers.
T-Trans and leadership shift (which will be described below)
are employed in the CTLCE framework, which is referred to as
the \emph{locally centralized T-Trans}~(LCTT), hereafter.
\par

Fig.~\ref{fig:Team-Trans} shows the T-Trans structure.
First, agent $i$'s observation 
is organized as a collection of information corresponding to each entity within its observational range: $o_i = (o_{i,1},\dots,o_{i,i},\dots,o_{i,|E|})$, and
$o_{i,j}=\emptyset$ for any entity that falls outside of 
$i$'s observable area. Subsequently, 
they are passed to a \emph{gate recurrent
  unit}~(GRU)~\cite{cho2014learning} cell~(see (a) in
Fig.~\ref{fig:Team-Trans}) to obtain temporal information. Then they
are passed to an attention-like module~\cite{vaswani2017attention}
(see (b) in Fig.~\ref{fig:Team-Trans}), to
represent the associations among entities. 
Agent's characteristics and the temporal features are
concatenated, and then they forwarded through an action head for calculating
Q-values. Additionally,
a classifier, which is shown as (c) in Fig.~\ref{fig:Team-Trans}, is
applied to determine the most 
appropriate candidate among the agents as the leader for the next time.
\par

In the attention-like module,
the calculation of the attention feature is executed as
\begin{equation} \label{eq:attention}
  \bm{z_{i}}= \textrm{softmax}(\frac{\bm{q_i} \bm{k_i}^T}{\sqrt{d_{att}}}) \bm{v_i},
\end{equation}
where $\bm{k_i}^T$ represents the transpose of $\bm{k_i}$;
$\bm{z_i}$ encapsulates a representation unique to
each teammate among the agents.
Then, $\bm{z_i}$ is concatenated with the duplicated 
hidden states and converted into teammate-specific
instructions by the fully connected (FC) layer action head.
\par

\Black{}

\subsection{Leadership Shift (LS)}\label{sec:LeadershipShift}
At the start of each episode, we randomly select 
$l$ ($>0$) agents as leaders~(Line~\ref{alg:LCE-l-leaders} in
Alg.~\ref{alg:LCE}).
The LS mechanism was introduced to facilitate the dynamic allocation of leadership and
to ensure that instructions are sent to workers from appropriate
leaders.
An MLP~((c) in Fig.~\ref{fig:Team-Trans}) is used by leaders
to generate the hidden states 
and calculate the {\em leadership scores}, 
$g_i=(g_{i,1},\dots,g_{i,n})$, 
where $g_{i,j}$ is the score of $i$'s assessment for $j$'s
potential as a leader.
Based on these scores, a leader nominates one of the agents it
observes to take over as a next leader. 
Algorithm~\ref{alg:LS} is the pseudocode of LS for agent $i$.
\par

To train the leadership scores 
$g_i$ effectively, we propose the
\emph{leadership Q-value}
to create pseudo-labels for $g_i$ as follows.
First, the leadership Q-value
for all agent $\forall k\in \AgentSet$ is calculated as
\begin{equation} \label{eq:LeaderQ}
  \bar{Q}^L_k(o_k, o_j, |\bar{\theta}) = \sum_j \bar{Q}_j
  (o_j, \textrm{argmax}_{a_j}{\bar{Q}_k(o_k, a_j | \bar{\theta})}
  |\bar{\theta}),
\end{equation}
where $j$ is an agent observable by $k$.
Note that $o_j$ is required to calculate 
$\bar{Q}^L_k$ only in the CT phase.
Our method in the LCE phase
does not depend on acquiring observations from 
other agents.
Then, the pseudo-labels for $g_i$, denoted by $g^*_i$,
are obtained using the onehot function:
\begin{equation} \label{eq:label-g}
  g^*_i = \textrm{onehot}(g_{i,1}^*,\dots,g_{i,n}^*).
\end{equation}
where $g_{i,k}^*=\bar{Q}^L_k$ if $i$ can observe $k$ and 
 $g_{i,k}^*= -\infty$, otherwise.
After that, $g_i$ is learned through the minimization of the
cross-entropy loss:
\begin{equation}\label{eq:loss-g}
  L_i^g(\theta) 
  = H(g'^*_i) 
  + D_{KL}(g'^*_i||g_i),
\end{equation}
where $D_{KL}(g'^*_i||g_i)$ is the Kullback-Leibler 
divergence of $g'^*_i$ from $g_i$, and $H(g'^*_i)$ is the entropy of
$g'^*_i$.
Note that $g'^*_i$ is obtained from the observations
at $t+1$ as the output of the target network with $\bar{\theta}$,
whereas $g_i$ is computed using the observations of $i$ at $t$ from
the network specified by $\theta$.
\par

Finally, the Q-values generated by the agents
are fed into a {\em mixing network}~\cite{rashid2020monotonic}
to combine these values effectively, 
and the collective learning goal for all parameters is
\begin{equation}\label{eq:loss-CTLCE}
  L(\theta) 
  = \sum_{i=1}^n L^Q_i(\theta)  
  + \lambda \sum_{i=1}^n L^g_i(\theta),
\end{equation}
where $\lambda$ is a weight hyperparameter and
$L^Q_i(\theta)$ is the standard 
DQN loss function~(Eq.~\ref{eq:loss-DQN}).
\par

\begin{algorithm}[tb]
\caption{Leadership Shift (LS)}
\label{alg:LS}
\begin{algorithmic}[1] 
\STATE Calculate the leadership scores $g_i=(g_{i,1},\dots,g_{i,n})$.
\STATE Choose the agent $j$ that is best suited to be the leader 
based on $j=\textrm{argmax}_j g_{i,j}$.
\IF{$j$ is not appointed to be a leader at $t+1$}
  \STATE Send a signal to $j$ to 
    appoint it to be a leader at $t+1$.
  \STATE Appoint $i$ to be a \emph{worker} at the $t+1$.
\ELSE
  \STATE Appoint $i$ to be a leader at $t+1$.
\ENDIF
\end{algorithmic}
\end{algorithm}

\section{Experiments}
We conducted an experiment to evaluate our
method, LCTT, using the LBF environment, as shown in
Fig.~\ref{fig:LBF}, where the size of the grid world is $10 \times
10$, the number of agents $n=4$ and the number of 
food $\beta=2$. 
These results are then compared to those with the baselines, 
QMIX~\cite{rashid2020monotonic} and MAIC~\cite{yuan2022multi}.

The level of any agent, $lv_a(i)$ is set to an integer randomly
selected between $1$ and $5$.
The level of food $f_\mu$ is also set tp a random integer 
$1\leq lv_f(\mu) < \sum_{i\in \AgentSet}\agentlv(i)$.
The visibility for all agents was confined to a $5\times 5$ area.
The setup for our experiments matched the conditions found in the
official MAIC repository. 
\par

Fig.~\ref{fig:results-LBF} shows the experimental results,
in which LCTT-$l$L means the LCTT with $l$ leaders; for instance,
LCTT-1L corresponds to a scenario where one leader
instructs others although the leadership can shift among agents. 
LCTT-0L correspond to a DE setup, 
while LCTT-4L depicts a scenario where every agent attempts to
instruct other agents.
\par

\begin{figure}[tb]
\centering
\includegraphics[width=0.96\linewidth]{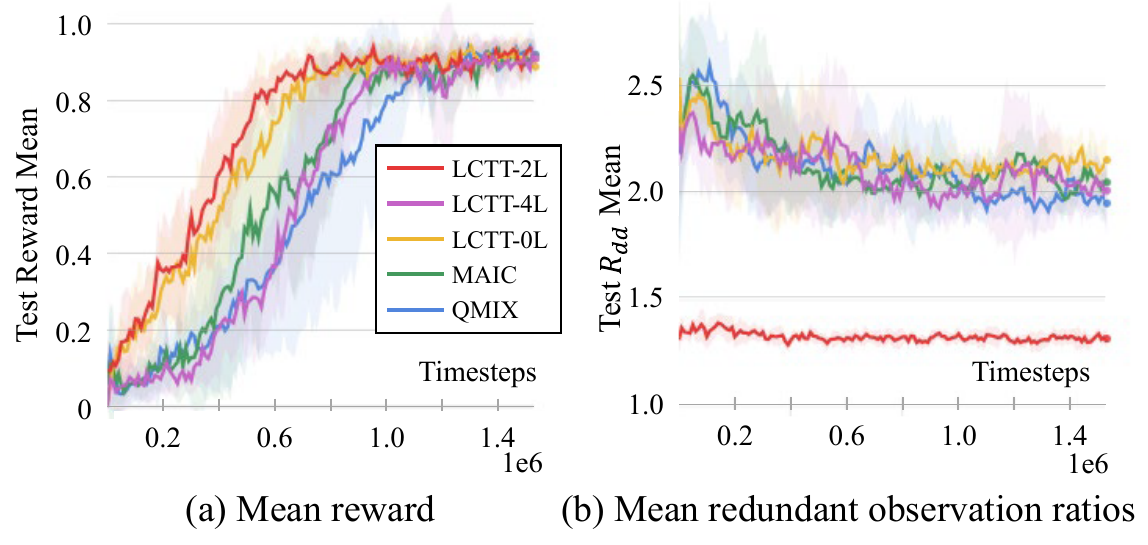}
\caption{Experimental results of LCTT and baselines.}\label{fig:results-LBF}
\label{fig:results-LBF}
\end{figure}

Figure~\ref{fig:results-LBF}a displays the mean rewards obtained by LCTT 
and baseline methods over timestep in the test phase.
While all approaches eventually reached comparable reward upon 
convergence, the LCTT with two leaders, denoted as LCTT-2L, 
achieved the fastest convergence.
MAIC enables each agent to send incentive messages that can bias 
the value functions of other agents.
In this case, MAIC can be considered a special case of LCTT in which $l=n=4$.
The faster convergence of LCTT-4L compared to MAIC could be attributed to the utilization of the T-Trans structure for message generation, 
rather than relying on teammate modeling. This is because T-Trans bypasses the need for explicit modeling or an ancillary loss function.
QMIX is a classical CTDE method, underpinning both LCTT and MAIC,
where agents decide based solely on their observations
without communication. 
From this perspective, QMIX represents a version of LCTT with $l=0$,
i.e., all agents are workers and make their own decisions,
without using T-Trans.
The founding that LCTT-0L converged faster than QMIX
suggests that the T-Trans framework is more effective than conventional MLP networks in learning.
For clarity, the performance curves of LCTT-1L and LCTT-3L were excluded from Fig.~\ref{fig:results-LBF}. They both converged faster than MAIC yet not as 
fast as LCTT-2L.
\par

Figure~\ref{fig:results-LBF}b presents the mean values of redundant observation ratio, $\redc$. 
It's clear that LCTT-2L has a significantly lower redundant observation
ratio compared to other methods, suggesting that two leaders of LCTT-2L effectively instruct worker agents in collaborative
tasks rather than let workers decide their own actions.
For LCTT-4L, LCTT-0L, MAIC, and QMIX, where all agents are required to make decisions, the redundant observation ratios are higher than those seen 
in LCTT-2L.
Moreover, their redundant observation ratios
temporarily increased at the beginning of the learning process,reflecting the agents' realization that solo efforts are insufficient for task completion, necessitating collective action. 
Over time, these ratios then level off to a more stable range as agents learn the importance of spreading out to effectively scout for food sources.
Unlike the others, the redundant observation ratio for LCTT-2L remains low throughout, only slightly decreasing as learning progresses, benefiting from the LCE strategy.
This strategy ensures that while agents come together, the 
redundant observation ratio does not rise because workers fall within the leaders' instruction region and cease their individual observations; 
conversely, as agents spread out, the 
redundant observation ratio does not fall further since workers exit the leaders' instruction region and resume their observation.
\par

\section{Conclusion}
This research addresses the issue of redundant computation in multi-agent systems. 
We introduced a novel metric, the redundant observation ratio, to quantitatively assess the extent of redundant computations.
Our findings suggest that reducing redundancy is feasible through a structured system of instruction construction among agents.
We developed the LCTT framework, enabling agents to decide which
agents should instruct and be instructed by others, and how.
Through the experimental results in LBF,
we demonstrated that the proposed method 
significantly reduces redundant computation without
causing a reduction in rewards, but instead achieving faster convergence.
Thus, when our method is applied to practice, it helps to save a large computational cost.
\par

We believe that it is promising to 
combine our study with 
research on multi-agent communication, for example, establishing
communication among leaders pave the way for even more significant advancements.
\par

\bibliography{lctt}
\end{document}